\documentclass[sigconf]{acmart}
\usepackage{balance}
\usepackage{multicol}
\usepackage[frozencache]{minted}
\usepackage{enumitem}
\usepackage{subcaption}
\usepackage{cprotect}
\usepackage{fancyvrb}
\VerbatimFootnotes
\DeclareMathOperator*{\argmin}{\mathsf{argmin}}

\AtBeginDocument{%
  \providecommand\BibTeX{{%
    \normalfont B\kern-0.5em{\scshape i\kern-0.25em b}\kern-0.8em\TeX}}}

\acmSubmissionID{40}

\copyrightyear{2021}
\acmYear{2021}
\setcopyright{acmcopyright}\acmConference[SIGCSE '21]{Proceedings of the 52nd ACM Technical Symposium on Computer Science Education}{March 13--20, 2021}{Virtual Event, USA}
\acmBooktitle{Proceedings of the 52nd ACM Technical Symposium on Computer Science Education (SIGCSE '21), March 13--20, 2021, Virtual Event, USA} \acmPrice{15.00}
\acmDOI{10.1145/3408877.3432358}
\acmISBN{978-1-4503-8062-1/21/03}

\begin{document}
\fancyhead{}

\title{Automating Program Structure Classification}

\author{Will Crichton}
\email{wcrichto@cs.stanford.edu}
\affiliation{%
  \institution{Stanford University}
}

\author{Georgia Gabriela Sampaio}
\email{gsamp@stanford.edu}
\affiliation{%
  \institution{Stanford University}
}

\author{Pat Hanrahan}
\affiliation{%
  \institution{Stanford University}
}

\renewcommand{\shortauthors}{Crichton et al.}

\begin{abstract}
  When students write programs, their program structure provides insight into their learning process. However, analyzing program structure by hand is time-consuming, and teachers need better tools for computer-assisted exploration of student solutions. As a first step towards an education-oriented program analysis toolkit, we show how supervised machine learning methods can automatically classify student programs into a predetermined set of high-level structures. We evaluate two models on classifying student solutions to the Rainfall problem: a nearest-neighbors classifier using syntax tree edit distance and a recurrent neural network. We demonstrate that these models can achieve 91\% classification accuracy when trained on 108 programs. We further explore the generality, trade-offs, and failure cases of each model.
\end{abstract}




\newcommand{\todo}[1]{\textcolor{red}{#1}}

\maketitle

\section{Introduction}
When a teacher creates a new programming assignment, they often wonder: what different kinds of solutions did my students come up with, and why? The strategies that students use, the way they organize their code --- this \textit{program structure} can reveal what students have (or haven't) learned. Classifying the structure of student solutions has been used to identify misconceptions\,\cite{wu2019zero}, success predictors\,\cite{wang2017learning}, and problem solving milestones\,\cite{yan2019pyramidsnapshot}. Studies on plan composition --- how students combine code templates to solve programming problems --- have long used program structure to analyze student problem solving. For example, Fisler's 2014 study of plan composition in functional languages showed that certain high-level program structures correlated with increased error rates\,\cite{fisler2014recurring}.

However, analyzing program structure is a challenging, time intensive work. In personal correspondence, Fisler estimated that hand-coding program structure for her 2014 study took 1-2 minutes per program. This estimate is consistent with Wu et al.\ who reported that hand-labeling student misconceptions in Code.org programs took an average of 2 minutes per program\,\cite{wu2019zero}. For CS1 courses with hundreds of students, such a per-program cost is prohibitive, motivating the use of automation to alleviate the burden of manual inspection.

Our vision for the future is that automatic program structure classification should be a tool in every CS teacher's toolbox. This analysis augments traditional forms of feedback (grades, office hours, etc.) with a new channel for understanding how students approach problems. With this tool, a teacher could explore the structural variation in student programs, assisted by data-driven technology to avoid a purely manual inspection process.

However such a tool does not exist today, so our goal is to lay the foundations for its development. That goal starts with the questions: what technologies could be used, and how well do they work? Ethical use of algorithms to analyze students requires a deep understanding of their accuracy and failure modes. Hence, in this paper, we perform a thorough evaluation of multiple tools for classifying student program structure. We address the following research questions:

\vspace{0.5em}\noindent \textbf{RQ1.} How much training does this tool need to be accurate? \\ 
\textbf{RQ2.} How accurate is the tool on different languages? \\
\textbf{RQ3.} When and why does this tool make errors? \vspace{0.5em}

\noindent We answer these questions for a well-studied programming problem: Rainfall, a simple list-processing task. We evaluate two supervised machine learning methods, nearest-neighbors and recurrent neural networks, on an existing dataset of student solutions to Rainfall in the OCaml and Pyret languages. We demonstrate that these approaches can classify Rainfall program structures with up to 91\% accuracy when trained on 108 examples.

\section{Related Work}

\begin{figure*}[t!]
    \centering
    \begin{subfigure}[t]{0.45\textwidth}
    \begin{minted}[linenos]{ocaml}
let rec help (alon : float list) = 
  match alon with
  | [] -> []
  | hd::tl ->
    if hd = (-999.) then []
    else if hd >= 0. then hd :: (help tl) 
    else help tl : float list)
let rec rainfall (alon : float list) =
  match alon with
  | [] -> failwith "no proper rainfall amounts"
  | _::_ ->
    (List.fold_right (+.) (help alon) 0.) /.
      (float_of_int (List.length (help alon))
\end{minted}
    \end{subfigure}%
    ~
    \begin{subfigure}[t]{0.45\textwidth}
    \begin{minted}[linenos]{ocaml}
let rec cut_list (a : int list) =
  match a with
  | [] -> []
  | hd::tl -> 
    if hd <> (-999) then hd :: (cut_list tl) else [] 
let non_neg (a : int list) = 
  List.filter (fun x -> x > 0) a
let average (a : int list) =
  (List.fold_right (fun x -> fun _val -> x + _val) a 0) / 
    (List.length a)
let rainfall (a : int list) = 
  average (non_neg (cut_list a))
\end{minted}
    \end{subfigure}

    \caption{Two example ``Clean First'' OCaml programs. Even for the same high-level structure, student solutions exhibit a significant diversity in syntactic and semantic variation such as the function decomposition strategies shown here.}
    \label{fig:example_ocaml_programs}
\end{figure*}

\subsection{Program classification}

Many kinds of high-level program analysis can be viewed as program classification. Plagiarism detection systems like MOSS\,\cite{bowyer1999experience} take a given student's program, and classify other programs as ``plagiarized'' or ``different''. We do not consider this task as program \textit{structure} classification, as plagiarism systems predominantly compare low-level syntax differences, e.g. whether two programs are the same modulo renamed variables.

Other prior works attempt to classify the kind of problem being addressed in a program, or what algorithm is being used. Taherkhani and Malmi\,\cite{taherkhani2013beacon} classify sorting algorithms from source code using decision-trees on hand-engineered problem-specific code features. For example, their features included ``whether or not the algorithm implementation uses extra memory'' and ``whether from the two nested loops used in the implementation of the algorithm, the outer loop is incrementing and the inner decrementing.'' 
Moving away from hand-engineering program features, the software engineering community has also applied deep learning techniques for similar tasks.  Mou et al.\,\cite{mou2016convolutional} defined a task of classifying which of 104 programming competition problems a program is attempting to solve, and they apply a novel tree-based neural network for this task. Bui et al.\,\cite{dq2019bilateral} introduce bilateral neural networks to solve the same problem in a language-independent manner. In this work, we focus on classifying how a student solved a problem, as opposed to what problem they were solving.

Closer to our application domain, Ahmed and Sindhgatta et al.\,\cite{tegcer2019} use program structure classification to suggest repairs to students. Wu et al.\,\cite{wu2019zero} evaluate a recurrent neural network (RNN) and multimodal variational autoencoder on classifying misconceptions in Code.org programs. In their problem formulation, a student can have one of 20 misconceptions about geometric concepts, and the goal is to classify which misconceptions a student has from their program. Malik and Wu et al.\,\cite{malik2019generative} introduce a method for neural approximate parsing of probabilistic grammars, achieving human-level accuracy on misconception classification. We adapt their RNN model in Sections \ref{subsec:rnn}.

\subsection{Program clustering and similarity}
\label{sec:related_cluster}

In a classification problem, a fixed set of categories is given up front. The role of a model is to classify data into one of these predetermined categories. Clustering methods attempt to solve a more challenging problem by simultaneously discovering the categories and the mapping from data to category. In the education community, prior work in program clustering has used classical heuristics such as computing edit distance between abstract syntax trees\,\cite{huang2013syntactic} and control-flow ASTs\,\cite{hovemeyer2016control}, or finding exact matches on canonicalized source code\,\cite{glassman2015overcode}, control-flow graphs\,\cite{luxton2013differences}, or simulation relations \cite{clara2018automated}. Edit distance has also been applied to program repair\,\cite{bader2019getafix} and code clone detection\,\cite{wang2018ccaligner}. 

We do not attempt to solve the problem of discovering program categories. In practice, one of the major challenges for clustering-based analyses is that the generated categories can range from hard to interpret to nonsensical. In our problem setting, we assume a teacher or CSE researcher has already identified categories of interest, and wants to label them at scale on a dataset of programs. However, we do use the notion of program distance via syntax-tree edits for our nearest-neighbors classifier.

Recent work has also applied machine learning techniques to learn program comparison metrics from data. Tufano et al.\,\cite{tufano2018deep} use a recursive autoencoder on identifiers, syntax trees, control flow graphs, and bytecode to build a semantic embedding space for programs, then use embedding space distance for clone detection.  Raychev et al.\ used decision trees\,\cite{raychev2016probabilistic} and conditional random fields\,\cite{raychev2015predicting} to learn associations between code fragments for predicting the values and types of holes in programs. While these metrics are likely more robust than tree edit distance, they are challenging to adapt for niche teaching languages like Pyret. For example, Raychev et al.\ used 150,000 JavaScript files to learn an embedding for JavaScript, and we strongly suspect we cannot find that much Pyret code out in the world.
\section{Dataset}
\label{sec:dataset}

Our goal is to evaluate program structure classification methods on programs and classes relevant to CSE researchers, i.e. classifying strategies on student programs, not classifying the kind of problem solved in LeetCode solutions. We chose to replicate the hand-labeled program structures used in Fisler's study of plan composition in functional solutions to the Rainfall problem\,\cite{fisler2014recurring}. In that study, students were prompted with:
\begin{quote}
    Design a program called rainfall that consumes a list of numbers representing daily rainfall amounts as entered by a user. The list may contain the number -999 indicating the end of the data of interest. Produce the average of the non-negative values in the list up to the first -999 (if it shows up). There may be negative numbers other than -999 in the list.
\end{quote} 

\noindent Fisler's dataset contains student solutions from different introductory programming classes in three functional languages: OCaml, Pyret, and Racket\footnote{We exclude Racket from our analysis because Fisler's Racket data were PDFs of hand-written exam solutions, which we cannot automatically analyze like a text file.}. Across these languages, Fisler identified  three high-level structures (``Single Loop'', ``Clean First'', and ``Clean Multiple'') that accounted for a large majority of student solutions. Each category indicates a different choice of when and how to filter the input list for valid rainfall data. Single Loop fuses summing/counting with filtering, Clean First filters the list then sums and counts the clean data, and Clean Multiple separately filters in the summing and counting logic. Figure~\ref{fig:example_ocaml_programs} shows examples of two Clean First OCaml programs, and Section~4 of Fisler's paper contains further discussion.

Overall, the dataset consists of 136 OCaml and 42 Pyret student solutions to the Rainfall problem. The distribution of Clean First / Clean Multiple / Single Loop solutions is .44/.20/.36 in OCaml and .47/.22/.31 in Pyret. Each solution's structure has been hand-labeled by a human expert (either Fisler or the current authors). In Section~\ref{sec:method}, we describe the methods for automatically classifying Rainfall program structures, and in Section~\ref{sec:eval} we evaluate the methods on Fisler's dataset.
\section{Methods}
\label{sec:method}

We selected which methods to evaluate based on several criteria:
\begin{itemize}[leftmargin=*]
    \item \textbf{Generality}: we prefer methods that could work with little customization for many languages and problems. So we eliminated any heuristic-based methods (e.g. as in Taherkhani and Malmi\,\cite{taherkhani2013beacon}), and only considered methods that learn directly from data.
    \item \textbf{Interpretability}: machine learning methods often trade off interpretability for predictive power. Neural networks are noted for their high accuracy and black-box nature, so we wanted to include classical machine learning methods as well.
    \item \textbf{Supervision}: as mentioned in Section~\ref{sec:related_cluster}, we only want to consider methods that learn provided categories, not emergent ones from the data. So we eliminate any unsupervised machine learning methods from consideration.
\end{itemize}

\noindent Given these criteria, we selected two methods: a nearest-neighbors classifier, and a recurrent \noindent neural network. We will explain each in greater detail:

\subsection{Nearest-neighbors classifier}
\newcommand{\dataset}{\mathcal{D}}
A nearest-neighbors classifier represents a baseline for supervised program structure classification. It has a simple formulation, requires no training algorithm, and makes interpretable decisions. To explain, let's set up the mathematical structure of the problem. The input is a dataset $\dataset = \{(p_1, l_1), \ldots, (p_n, l_n)\}$ of $n$ pairs of programs $p$ and labels $l$. For example, the two programs in Figure~\ref{fig:example_ocaml_programs} both have $l = \text{"Clean First"}$.

Given a new program $p'$, a nearest-neighbors classifier finds the most similar program $p_i$ from the training data, and assigns $p'$ the label $l_i$. Formally, given a distance function:
$$\mathsf{Dist} : \mathsf{Program} \times \mathsf{Program} \rightarrow \mathbb{R}$$ A nearest-neighbors classifier classifies a new program $p'$ as:
$$\argmin_{p, l \in \mathcal{D}} \mathsf{Dist}(p, p')$$
This method is interpretable in that the classifier can provide the nearest training program $p$ as a justification for its classification. See Figure~\ref{fig:incorrect_plans} for an example.

The key design decision is choosing a distance metric. One metric that has been widely used for program similarity is \textit{tree edit distance}. When two programs are represented as their abstract syntax tree (AST), the edit distance is the number of tree manipulation operations needed to transform one tree into the other. For our classifier, we use the canonical Zhang-Shasha method\,\cite{zhang1989simple}. Nearest-neighbors could, of course, be used with other distance metrics (e.g. Euclidean distance in a learned embedding space). But for simplicity in this paper, we will use ``nearest-neighbors'' to mean ``with Zhang-Shaha distance.''

Similar to prior work\,\cite{hovemeyer2016control}, we do not compare syntax trees verbatim. Small syntactic differences like choice of variable name or presence of type annotation do not usually impact the high-level structure of a program. For both OCaml and Pyret, we use their respective compilers to generate a raw AST, then erase variable names, constant values, and type annotations before computing edit distance.

\subsection{Recurrent neural network}
\label{subsec:rnn}

\begin{figure}[t]
    \centering
    \includegraphics[width=\columnwidth]{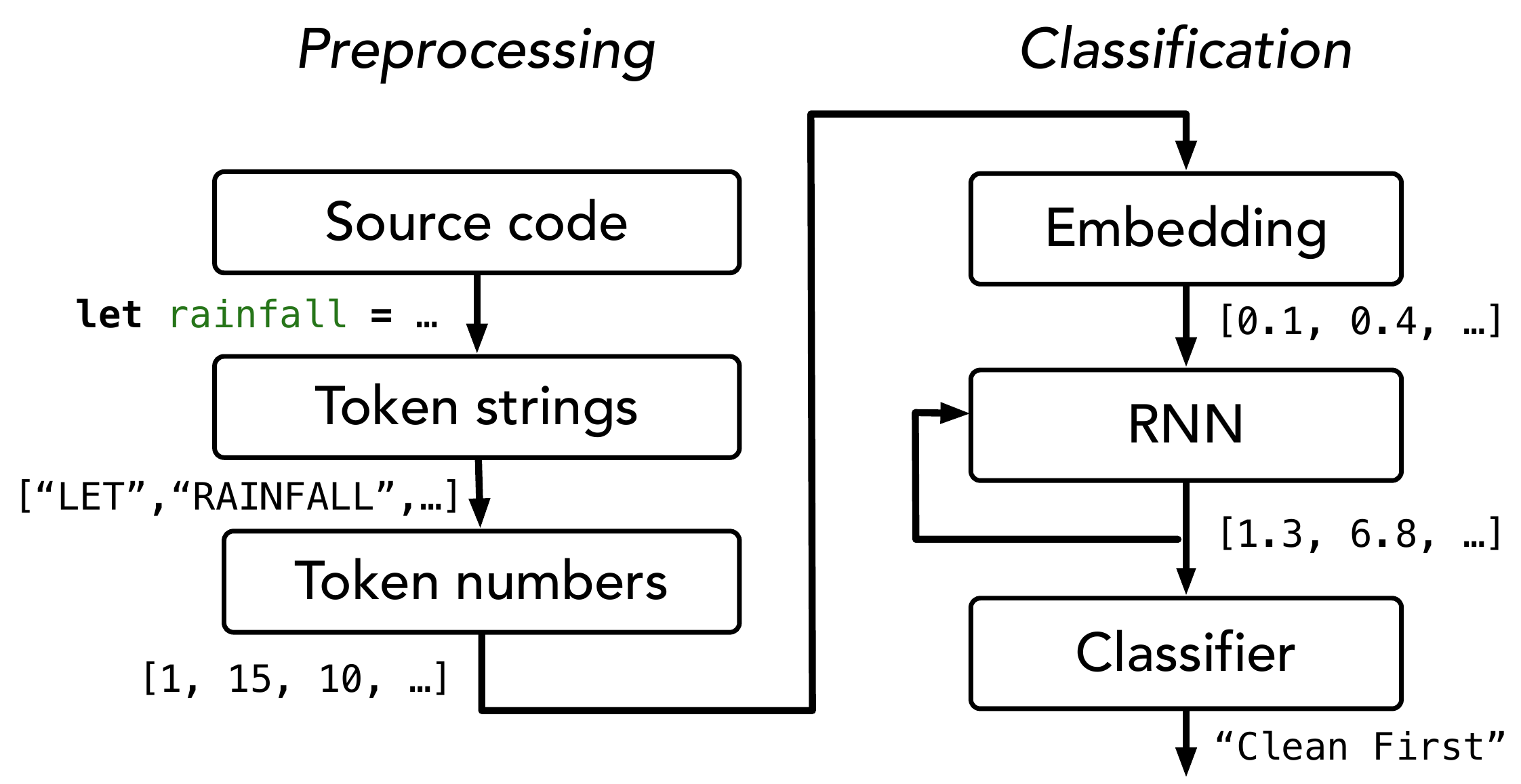}
    \caption{RNN classifier architecture. Programs are converted into integer sequences where each number uniquely identifies a token. Each token is mapped to an embedding vector, fed through the RNN to produce a hidden state vector. At the end of the sequence, a classifier predicts the program category from the hidden state. \vspace{-1em}}
    \label{fig:rnn_architecture}
\end{figure}

A downside to nearest-neighbors is that the distance metric is susceptible to issues where programs may be syntactically similar but structurally different (or vice versa, as in Figure~\ref{fig:example_ocaml_programs}). Neural networks are a supervised learning method that automatically learn program similarity features from the training data. In practice, learned features can increase accuracy (with enough training data), but decrease interpretability. We use a \textit{recurrent} neural network because our input programs do not have a fixed size,\footnote{We cannot simply use a standard feed-forward neural network with a large input size, since that presumes we know the maximum program size at training time. A student could always produce a program larger than previously observed.} unlike e.g.\ convolutional neural networks for image classification.

We adapt a basic RNN architecture from Wu et al.\,\cite{wu2019zero} as shown in Figure~\ref{fig:rnn_architecture}. Each token of the source program is mapped to a high-dimensional vector of numbers (an ``embedding'').\footnote{Although a token is represented as a number, tokens are still categorical, not ordinal data. For example, if \verb|let| = 1, \verb|fun| = 2, \verb|end| = 10, the network shouldn't learn that ``\verb|let|'' and ``\verb|fun|'' are somehow more related than ``\verb|end|'' by virtue of being assigned closer identifiers.}. The RNN is initialized with a different high-dimensional vector of numbers (the ``hidden state''). Given a sequence of token embeddings, the RNN iterates through each token and updates the hidden state with information from the embedding. At the end of the sequence, logistic regression is used to classify the hidden state into a probability distribution over the possible program structure categories.

Given the breakneck pace of research on neural networks, there are inevitably countless variations on this architecture that could be applied to our problem. We used the most basic possible recurrent architecture over e.g. transformers\,\cite{vaswani2017attention,lachaux2020unsupervised} or tree-LSTMs\,\cite{tai2015improved} to reduce the number of confounding factors that influence accuracy. We consider the specific choice of RNN cell (LSTM\,\cite{hochreiter1997long} vs. GRU\,\cite{cho2014properties}), the number of layers within the RNN, and the embedding/hidden vector sizes as hyperparameters.

To train the RNN model on labeled student data, we use gradient descent with Adam\,\cite{kingma2014adam} ($\alpha = 0.001$, $\beta_1 = 0.9$, $\beta_2 = 0.999$). We use the model weights from the training iteration with highest accuracy on a validation set as the version for evaluation. We use Bayesian optimization\,\cite{bergstra2013hyperopt} to select the best hyperparameters (LSTM cell, 3 layers, 512 embedding size, 128 hidden size).

\begin{figure}[t!]
    \centering
    \includegraphics[width=\columnwidth]{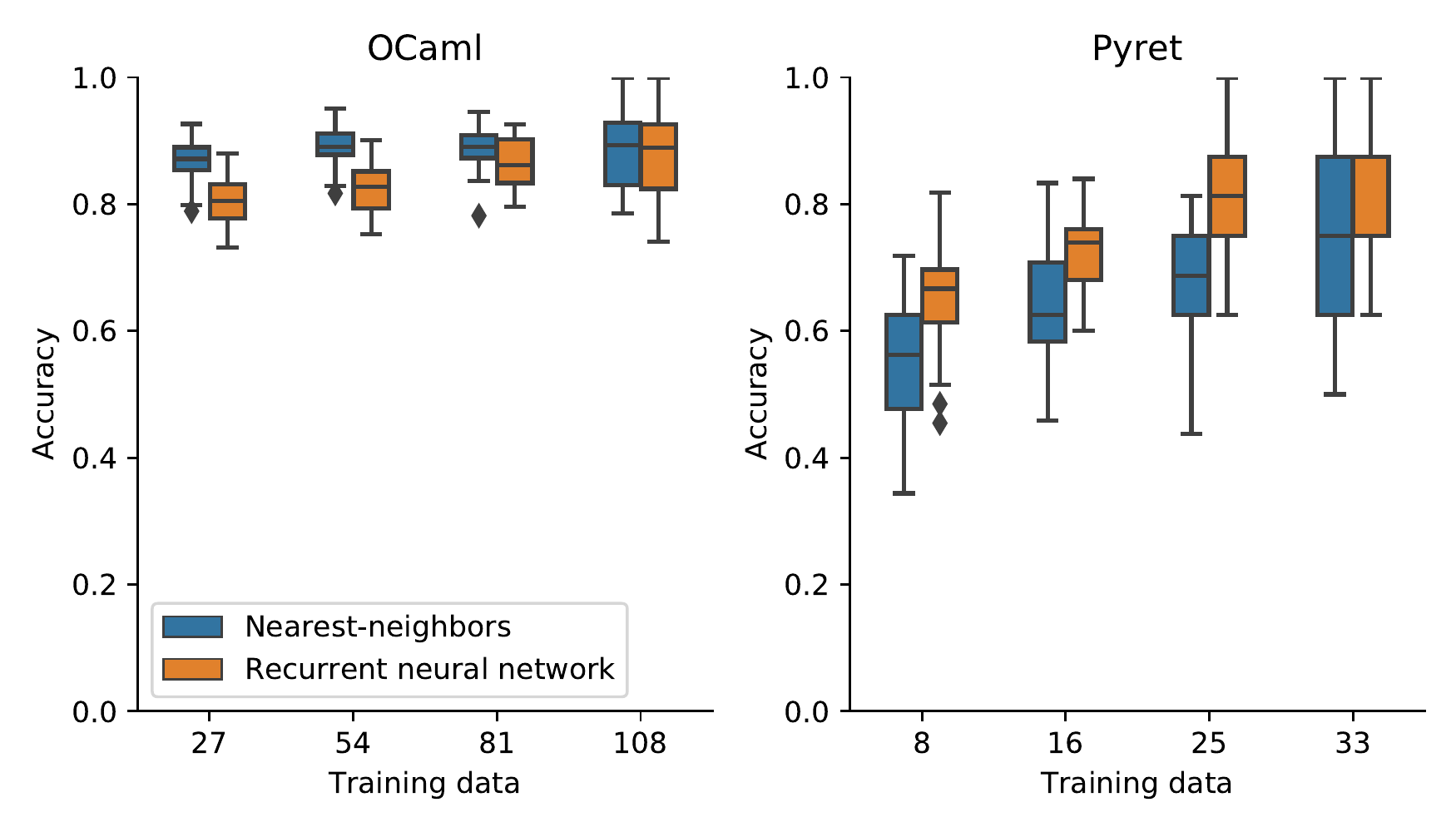}
    \caption{Distribution of model accuracy for different sizes of training and test sets and under each language. Each experimental condition is computed through 30 trials, so its distribution is visualized as box plot. \vspace{-1em}}
    \label{fig:student_data_comparison}
\end{figure}
\begin{figure}[t!]
    \centering
    \includegraphics[width=\columnwidth]{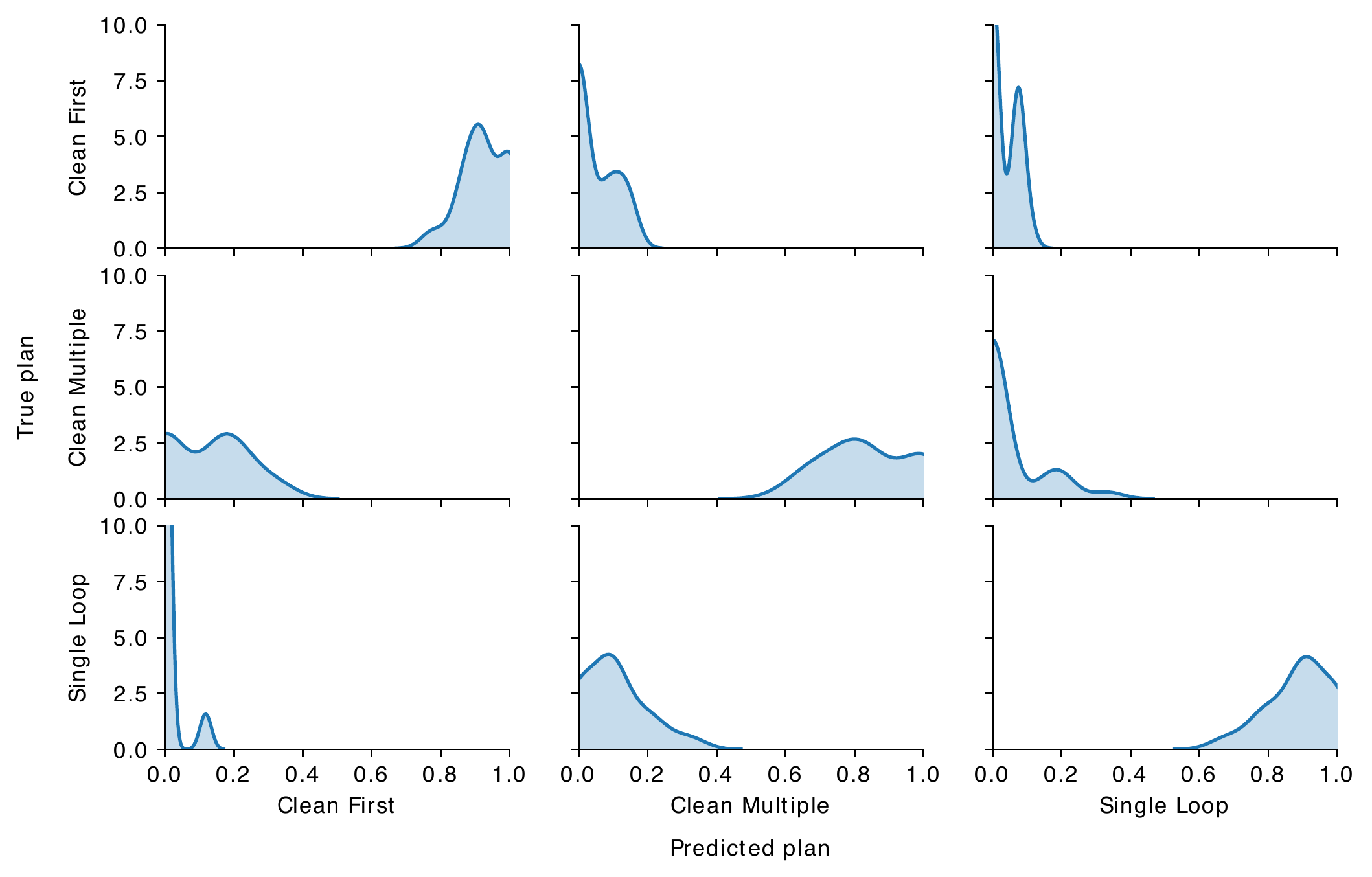}
    \caption{Distribution of row-normalized confusion matrices for the RNN method in OCaml. The $x$-axis of each sub-plot shows the probability $P(\text{Predicted plan} \mid \text{True plan})$, e.g.\ the upper right plot shows the probability a program is classified as Single Loop given its true plan is Clean First. The $y$-axis shows the number of times a confusion matrix had a given probability across the 30 simulations. Kernel density estimation is used to smooth the empirical probability distribution. The nearest-neighbors matrices look very similar, and so are not shown. \vspace{-1em}}
    \label{fig:confusion_matrix}
\end{figure}

\section{Evaluation}
\label{sec:eval}

For each method, we answer the three research questions raised in the introduction.

\subsection{How much training does this tool need to be accurate?}

To estimate the accuracy of a particular method, we can partition the Rainfall dataset into training data and testing data. The method is trained on the training data with knowledge of their ground-truth program structures. Then the method is evaluated on the testing data without knowing their ground-truth. For now, each test prediction is either correct if the predicted category matches the actual category, and incorrect otherwise. (Section~5.3 further distinguishes between types of errors.) The accuracy of the model is then the fraction of test programs with a correctly predicted category.

Each train/test partition acts as a simulation of how the tool would be used in practice. For example, if a teacher has 100 student solutions, they could hand-label 30 of them with the correct program structure, then train a model to predict the remaining 70. To understand the ``average'' scenario, we can run this simulation multiple times on random partitions to find a distribution of accuracy. Technically speaking, we use a Monte Carlo cross-validation.

To understand the relationship between accuracy and training data, we fix different amounts of training data as the independent variable, and measure the distribution of accuracy across 30 simulations as the dependent variable. Figure~\ref{fig:student_data_comparison} shows the distribution for each method over both languages at each quartile of training data. Some observations:

\begin{itemize}[leftmargin=*]
    \item The highest mean accuracy in OCaml is $.91 \pm .04$ for nearest-neighbors and $.89 \pm .05$ for RNN models when training on 108 programs. For Pyret, the highest mean is $.71 \pm .19$ for nearest-neighbors and $.87 \pm .10$ for RNN on 33 training programs.
    \item Nearest-neighbors achieves $.85 \pm .04$ mean accuracy for OCaml with only 27 training programs.
    \item Nearest-neighbors outperforms RNN under every training set size for OCaml, and the converse is true for Pyret. 
\end{itemize}

\begin{figure*}[t!]
    \centering
    \begin{subfigure}[t]{0.45\textwidth}
    \begin{minted}[linenos]{ocaml}
let rec rainfall_help1 (alon : int list) =
  match alon with
  | [] -> 0
  | (-999)::tl -> 0
  | hd::tl -> hd + (rainfall_help1 tl)
let rec rainfall_help2 (alon : int list) =
  match alon with
  | [] -> 0
  | (-999)::tl -> 0
  | hd::tl -> 1 + (rainfall_help2 tl)
let rainfall (alon : int list) =
  (float_of_int (rainfall_help1 alon)) /.
     (float_of_int (rainfall_help2 alon))
\end{minted}
    \end{subfigure}%
    ~
    \begin{subfigure}[t]{0.45\textwidth}
    \begin{minted}[linenos]{ocaml}
let rec positive lst =
  match lst with
  | [] -> []
  | head::tail ->
    (match head with
      | (-999) -> []
      | x when x > 0 -> head :: (positive tail)
      | x when x < 0 -> positive tail)
let rec sum_list lst =
  match lst with 
  | [] -> 0 
  | head::tail -> head + (sum_list tail)
let rec rainfall lst =
  (sum_list (positive lst)) / (List.length (positive lst))
\end{minted}
    \end{subfigure}
    \caption{The left program is a Clean Multiple solution that was misclassified by the nearest-neighbors classifier as Clean First, being matched with the Clean First program in the training set on the right. The functions share significant syntactic and structural similarity, e.g. two helper functions, similar style of match, and similar top-level usage. However, the helper functions are critically used in very different ways, leading to an incorrect classifier prediction. \vspace{-1em}}
    \label{fig:incorrect_plans}
\end{figure*}

\noindent For OCaml, relatively little training data is needed to achieve high accuracy with nearest-neighbors, while at least 100 data points is needed for RNN to achieve comparable accuracy. For Pyret, both methods have uncomfortably high variance even with 33 training programs, suggesting that more training data is needed for high, stable accuracy on Pyret programs with these methods.

\subsection{How accurate is the tool on different programming languages?}

\noindent We have to carefully compare the results of Figure~\ref{fig:student_data_comparison} between OCaml and Pyret due to the difference in dataset size. By comparing e.g. the 27-size OCaml condition vs. the 25-size Pyret condition, we can get close to a fair comparison. Observations:
\begin{itemize}[leftmargin=*]
    \item Nearest-neighbors gets $.85 \pm .04$ mean accuracy for OCaml, and $.67 \pm .08$ mean accuracy for Pyret.
    \item RNN gets $.79 \pm .04$ mean accuracy for OCaml, and $.79 \pm .08$ mean accuracy for Pyret.
\end{itemize}

\noindent For small amounts of data, the RNN performs consistently with about 79\% accuracy in both cases. In OCaml, nearest-neighbors outperforms the RNN with 85\% accuracy, but is substantially worse for Pyret with 67\% accuracy. This difference suggests that the RNN is more stably language-independent, while nearest-neighbors' performance is language-dependent.

For nearest-neighbors, the performance gap between the two languages is possibly explained by AST size. While the average program length in tokens is 116 for OCaml vs. 127 for Pyret, the average program size in number of AST nodes is 50 for OCaml and 196 for Pyret. Hence, the token-based RNN sees programs of much similar size than the node-based nearest-neighbors classifier. This difference is likely an artifact of the implementation of ASTs in the respective compilers. Further work in simplifying the AST could potentially improve nearest-neighbors performance.

\begin{table*}[th!]
\setlength{\tabcolsep}{0.5em} 
{\renewcommand{\arraystretch}{1.2}
\begin{tabular}{p{0.28\textwidth} | p{0.3\textwidth} | p{0.3\textwidth}}
 & \textbf{Nearest-neighbors} & \textbf{RNN} \\ \hline
\textbf{How much training does this tool need to be accurate?} & Few programs for small ASTs, many for large ASTs & More than nearest-neighbors to reach peak accuracy \\ \hline
\textbf{How accurate is the tool on different programming languages?} & Worse as AST size increases & Consistent across languages \\ \hline
\textbf{When and why does this tool make errors?} & Confuses syntactically similar and semantically different programs & Learns an embedding space that doesn't separate categories well enough
\end{tabular}}
\caption{Summary of answers to each research question for each method. \vspace{-2em}}
\label{tab:rqs}
\end{table*}

\subsection{When and why does this tool make errors?}

\noindent First, to understand statistically when errors are most likely occur, we ran 30 simulations of both models for the OCaml dataset in a 70/30 train/test split. Rather than evaluating accuracy (is the prediction correct or not?), we consider the more granular statistic of a confusion matrix: for each category, how often is it classified as a different category? Each simulation generates one confusion matrix, which we summarize as a distribution over matrices in Figure~\ref{fig:confusion_matrix}.

The plot shows that true Clean First programs are misclassified less often than the other two classes. Clean Multiple has a greater misclassification rate, being most frequently confused with Clean First. And Single Loop is almost exclusively misclassified as Clean Multiple, a somewhat confusing asymmetry given Clean Multiple is rarely misclassified as Single Loop. These observations are consistent between both nearest-neighbors and RNN.

Next, to understand the interpretability of these errors, we will answer a particular ``why'' question: why is Clean Multiple often misclassified as Clean First? Starting with nearest-neighbors: recall that programs are classified by their edit-distance to the nearest program in the training set. Given an incorrectly classified program, we can look at the closest training program to understand why the error occured.

Figure~\ref{fig:incorrect_plans} shows a representative example of a Clean Multiple program misclassified as Clean First by nearest-neighbors. The two programs shared many syntactic features (multiple helper functions, use of standard library functions, similar matching structure), but were subtly distinct in how these pieces of code were used. Through manual inspection, we found most of the errors from nearest-neighbors were caused by such incidental syntactic similarities.

\begin{figure}[t!]
    \centering
    \includegraphics[width=\columnwidth]{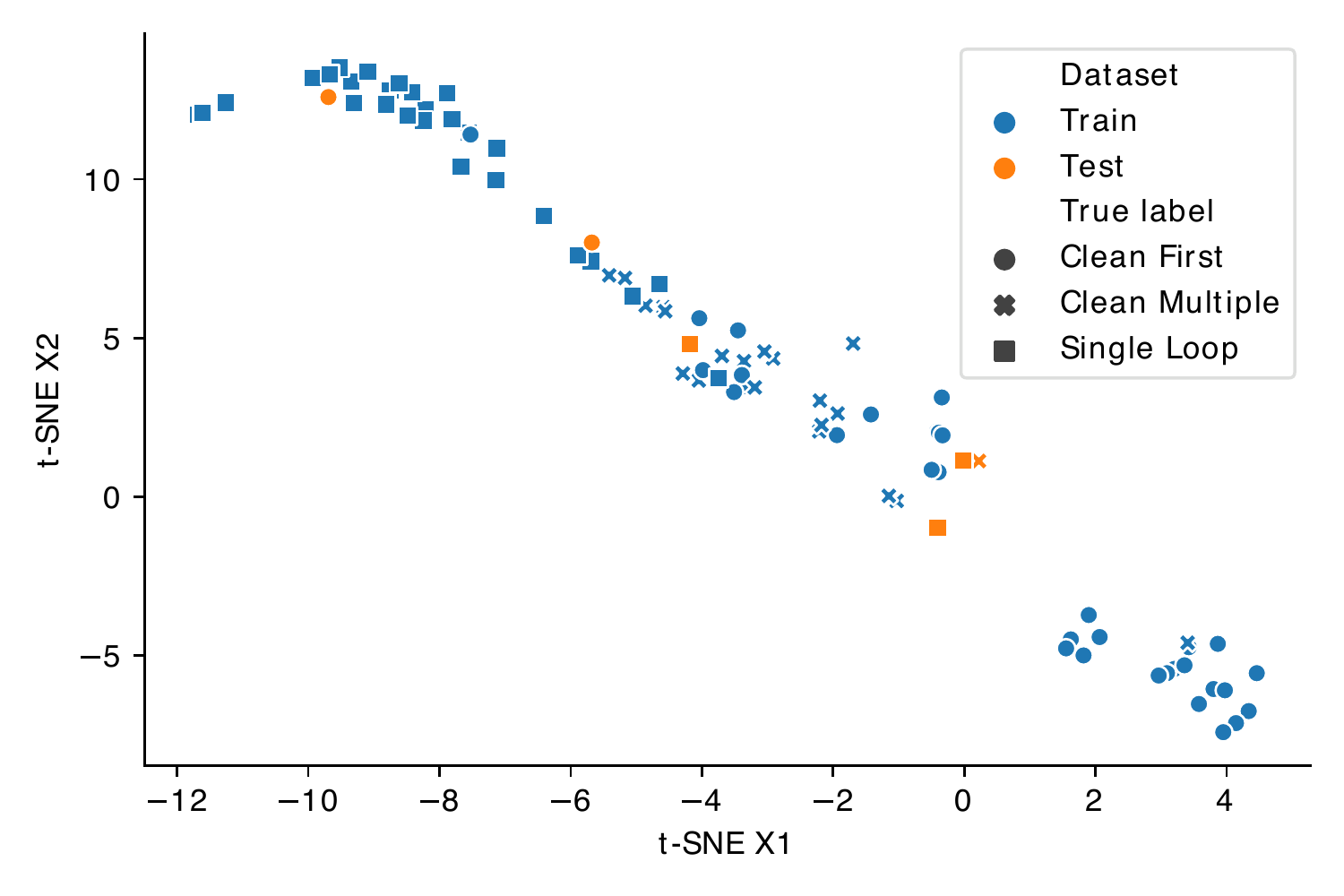}
    \caption{A scatter plot of the t-SNE projection of hidden state vectors for OCaml programs. Blue is training data, and orange is incorrectly classified test data. \vspace{-2em}}
    \label{fig:tsne}
\end{figure}

As with many neural-network approaches, the RNN provides no immediately human-interpretable ways to understand its predictions. However, research in interpretable machine learning has produced methods of visualizing the internal representations of objects within a neural network. Once a program has been fed to the RNN, it generates a "hidden state" vector of numbers. Using the t-SNE method\,\cite{maaten2008visualizing}, we can project that high-dimensional representation to a 2D plane as shown in Figure~\ref{fig:tsne}.

Given a particular random 70/30 split on the OCaml dataset, we generated a t-SNE diagram by projecting the 92 training programs onto a 2-D scatterplot, shown in blue. Then we add the 6 incorrectly classified test programs, shown in orange. Each point's shape indicates its actual category. The t-SNE diagram reveals that roughly three clusters emerge, one for each category: squares in the top-left (Single Loop), crosses in the middle (Clean Multiple), and circles in the bottom-right (Clean First). When a program is misclassified, its embedding is closer to the cluster of another category than its own.

For the question of Clean Multiple vs. Clean First: the t-SNE diagram shows that the RNN embedding space learns to position Clean Multiple programs \textit{between} Clean First and Single Loop. Hence, why Clean Multiple is disproportionately misclassified as one or the other, and why Single Loop is rarely misclassified as Clean First. Additionally, the one misclassified Clean Multiple program (the orange cross in Figure~\ref{fig:tsne}) is closer to Clean First training points than to Clean Multiple training points, a likely explanation for its misclassification. In sum, the RNN is not learning an embedding space that keeps programs of each category sufficiently far apart.

\section{Discussion}

We summarize the answers to our research questions in Table~\ref{tab:rqs}. Overall, we have found that these methods have the potential to classify functional Rainfall programs with relatively high accuracy (90\%+) without huge amounts of training data. The methods work across languages and have varying levels of interpretability. 

\subsection{Threats to validity}

While we hope that these methods can be used by teachers for their own programming problems, this study's conclusions may not generalize beyond problems like Rainfall. For example:
\begin{itemize}[leftmargin=*]
    \item \textbf{Problem complexity:} Rainfall is a simple problem, whose solutions in OCaml and Pyret are 10-20 lines. Methods like nearest-neighbors may not scale to more complex problems.
    \item \textbf{Language/paradigm:} the dataset's programming languages were both functional. These methods may not perform as well on Java, Python, or other more standard CS1 languages. 
    \item \textbf{Number of categories:} with more program structures, more data is needed to distinguish between them.  
\end{itemize}

\noindent As more datasets become available with CSE-relevant program classification tasks, we hope that these concerns can be addressed in future work. Additionally, as more powerful machine learning methods are developed, they can be applied to overcome the limitations of the methods evaluated here.

\subsection{Applications to CS education}

Thus far, program classification tools have primarily been the domain of CS education researchers with machine learning expertise. We hope that tools with UIs designed for ease-of-use and transparency/debuggability will make this technology accessible to all CS educators. For starters, all code from this project is free and open-source at \url{https://github.com/willcrichton/autoplan}. We have developed a simple Python API to simplify data preprocessing and model training specifically for program classification of Pyret, OCaml, Java and Python programs.


The bigger question is: how would teachers use such a tool? We expect that teachers could use program classification to gain visibility into the strategies used by students without needing to read every program. The average workflow might look like this:
\begin{enumerate}[leftmargin=*]
    \item A teacher notices in office hours that some students are writing their solutions a particular way, e.g.\ performing multiple validation checks up front vs.\ performing them lazily throughout the program (such as Clean First vs.\ Clean Multiple).
    \item After collecting assignment solutions, the teacher finds a few examples of solutions that do and don't match this pattern.
    \item The teacher trains a classification model on the examples, and uses it to find more similar programs to label.
    \item The teacher repeats this workflow until they are confident that the model is accurate for their dataset based on cross-validation simulations like those in this paper.
    \item They run the model on the entire solution dataset, revealing that 1/3 of the class is using the lazy validation strategy.
    \item The teacher prefers students to validate eagerly, and so updates their teaching materials to cover this issue in class.
\end{enumerate}

\noindent We hope that this study can contribute toward the foundational knowledge needed to make this process possible. 

\section{Acknowledgements}

We are deeply grateful to Kathi Fisler for providing us access to the Rainfall dataset.


\bibliographystyle{ACM-Reference-Format}
\bibliography{bibliography}

\end{document}